\documentclass[a4paper]{PoS}    

\usepackage{amsmath,amssymb}
\usepackage{physics}   

\usepackage{floatrow}    
\floatstyle{plaintop}   
\restylefloat{table}    
\usepackage{graphicx}
\usepackage{caption}   

\usepackage{cite}   

\title{Nucleon Magnetic Properties from Lattice QCD with the Background Field Method}

\ShortTitle{Nucleon Magnetic Properties from Lattice QCD with the Background Field Method}

\author{\speaker{Ryan Bignell}%
	\\	Special Research Centre for the Subatomic Structure of Matter, Department of Physics, University of Adelaide, South Australia 5005\\
	E-mail: \email{ryan.bignell@adelaide.edu.au}}

\author{Waseem Kamleh\\
        Special Research Centre for the Subatomic Structure of Matter, Department of Physics, University of Adelaide, South Australia 5005\\
}

\author{Derek Leinweber\\
	Special Research Centre for the Subatomic Structure of Matter, Department of Physics, University of Adelaide, South Australia 5005\\
	}

\author{Matthias Burkardt\\
	Department of Physics, New Mexico State University, Las Cruces, NM 88003-001, U.S.A.}

\abstract{The magnetic moment and magnetic polarisability of the neutron and proton are investigated using
	the uniform background-field method and lattice QCD. The results are calculated using $32^3\times64$
	dynamical QCD lattices provided by the PACS-CS collaboration through the ILDG. Methods of
	isolating the energy of the hadrons accounting for Landau level energies are explored. Effective energy
	shifts are formed from combinations of correlation functions and their efficiency in isolating the
	magnetic moment and polarisability terms is determined.}    

\FullConference{The 26th International Nuclear Physics Conference\\
	11-16 September, 2016\\
	Adelaide, Australia}

\begin{document}
	
	\section{Introduction}
	The magnetic properties of the nucleon such as the magnetic moment and polarisability can be accessed using lattice QCD with the background field method \cite{Martinelli:1982cb,Bernard:1982yu,Burkardt:1996vb,Lee:2005dq,Primer:2013pva,Chang:2015qxa}. The magnetic polarisability is a measure of the deformation of a system of charges in an external magnetic field. This deformation causes an energy shift in the particle which can be determined by the energy field relation \cite{Martinelli:1982cb,Tiburzi:2012ks}
	\begin{align}
		E(B) = M +\vec{\mu}\vdot\vec{B} + \frac{ \abs{qe\,B}}{2\,M} - \frac{4\,\pi}{2}\,\beta\,B^2 + \order{B^3}.
		\label{EofB}
	\end{align}
	Previous studies \cite{Primer:2013pva} have faced difficulty in extracting a reliable signal for the polarisability. This is due to the polarisability being a second order effect and the complication of the Landau levels which cannot be easily isolated from the polarisability.
	\par
	The Landau levels are a series of energy levels arising from a charge or system of charges in an external magnetic field. Hence the charged proton has a Landau level that must be accounted for. 
	In the absence of QCD, the consituent quarks would have individual Landau levels. It is an open question as to the extent to which this effect remains (if it all) in the presence of QCD interactions.	
	
	\section{Background Field Method}
	To introduce a background field on the lattice, first consider the continuum case. Here the covariant derivative is modified by the addition of an electromagnetic coupling
	\begin{align}
		{D}_\mu \rightarrow {D}_\mu^\prime = \partial_\mu +g\,G_\mu +qe\,A_\mu,
	\end{align}
	Where $qe$ is the charge on the fermion field and $A_\mu$ is the electromagnetic four-potential. Discretising this additional term in the same way as the usual gauge fields \cite{Gattringer:2010zz} results in the gauge links being multiplied by an exponential phase factor
	\begin{align}
		U_\mu(x) \rightarrow U_\mu(x)^{(B)} = \text{e}^{\,i\,a\,qe\,A_\mu(x)}\,U_\mu(x).
	\end{align}
	Thus far the electromagnetic gauge potential has not been specified uniquely. In order to obtain a magnetic field along the $\hat{z}$ axis, a potential
	\begin{align}
		A_x = -B\,\hat{y},
	\end{align}
	is used over the interior of the $N_x \times N_y \times N_z \times N_t$ lattice. The periodic boundary conditions of the lattice require a non-trivial potential to ensure that the field is uniform over the entirety of the lattice. This requirement produces a quantisation condition on the magnetic field strength \cite{Primer:2013pva}
	\begin{align}
		\abs{q_de\,B} = \frac{2\,\pi\,k_d}{N_x\,N_y\,a^2},
		\label{quantcond}
	\end{align}
	where $k_d$ is an integer governing the field strength.

	\section{Simulation Details}
	The calculations detailed here use $2+1$ flavour dynamical QCD configurations provided by the PACS-CS collaboration \cite{Aoki:2008sm} through the International Lattice Data Grid \cite{Beckett:2009cb}. These lattices have dimensions $32^3 \cross 64$ with $\beta = 1.9$ and a physical lattice spacing of $a = 0.0907(13)$ fm. A clover fermion action and Iwasaki gauge action are used. A single value of the light-quark hopping paramter, $k_{ud} = 0.13754$ corresponding to a pion mass of $m_\pi = 413$ MeV is used in this study. The lattice spacing for this mass was set using the Sommer scale with $r_0 = 0.49$ fm. The configuration ensemble size was 450.
	\par
	To be able to use Eq.~(\ref{EofB}) to extract the polarisabilities, correlation functions at four distinct magnetic field strengths are calculated. As the $u$ and $d$ quarks have different signs, separate propagators at different field strengths must be calculated for each distinct field strength. These correspond to $k_d = 0, \, \pm 1, \, \pm 2,\, \pm 3,\,\pm 4,\,\pm 6$ in Eq.~(\ref{quantcond}).
	\par
	The configurations used in this study did not include a background field when generated. Hence the only quarks which feel the presence of the external magnetic field are the valence quarks of the hadrons. To include the background field on the configurations requires separate gauge field configurations for each field strength. This is prohibitively expensive and also destroys the advantageous correlations between the field strengths.
	
	\section{Magnetic Polarisability}
	From correlation functions calculated with a background field in place, the magnetic polarisabilty can be extracted. To do this consider the energy-field relation in Eq.(~\ref{EofB}). We wish to remove the $\vec{\mu}\vdot\vec{B}$ and $M$ terms. This can be done by using the spin depdendence of the $\vec{\mu}\vdot\vec{B}$ term and the zero-field correlator. Taking a combination of spin orientations and field strengths produces the desired result for the energy shift
	\begin{align}
		\Delta E_p(B) &= \frac{1}{2}\, \left( E_{\uparrow}(B) + E_{\downarrow}(B) - E_{\uparrow}(0) - E_{\downarrow}(0) \right) \nonumber \\
					&= \frac{\abs{qe\,B}}{2\,M} - \frac{4\,\pi}{2}\,\beta\,B^2.
		\label{dE(B)}
	\end{align}
	A superior method with which to extract this energy is to take a ratio of the correlators directly. This has the advantage of allowing correlated errors to cancel prior to fitting
	\begin{align}
		R_p(B,t) =&  \left( \frac{G_{\downarrow}(B+,t) + 
			G_{\uparrow}(B-,t)  }{ G_{\downarrow}(0,t) + G_{\uparrow}(0,t)}
		\right) \, \left( \frac{G_{\downarrow}(B-,t) + G_{\uparrow}(B+,t) 
		}{G_{\downarrow}(0,t) + G_{\uparrow}(0,t)} \right).
	\end{align}
	Here the $\uparrow$ and $\downarrow$ represent spin up and down while $B\pm$ represents magnetic fields in the postive and negative $\hat{z}$ directions. From this ratio, an effective energy shift can be extracted in an analogous way to an effective mass.

	\section{Quark Projection}
\begin{figure}[t!]
		\begin{center}
			{\includegraphics[width=0.24\textwidth]{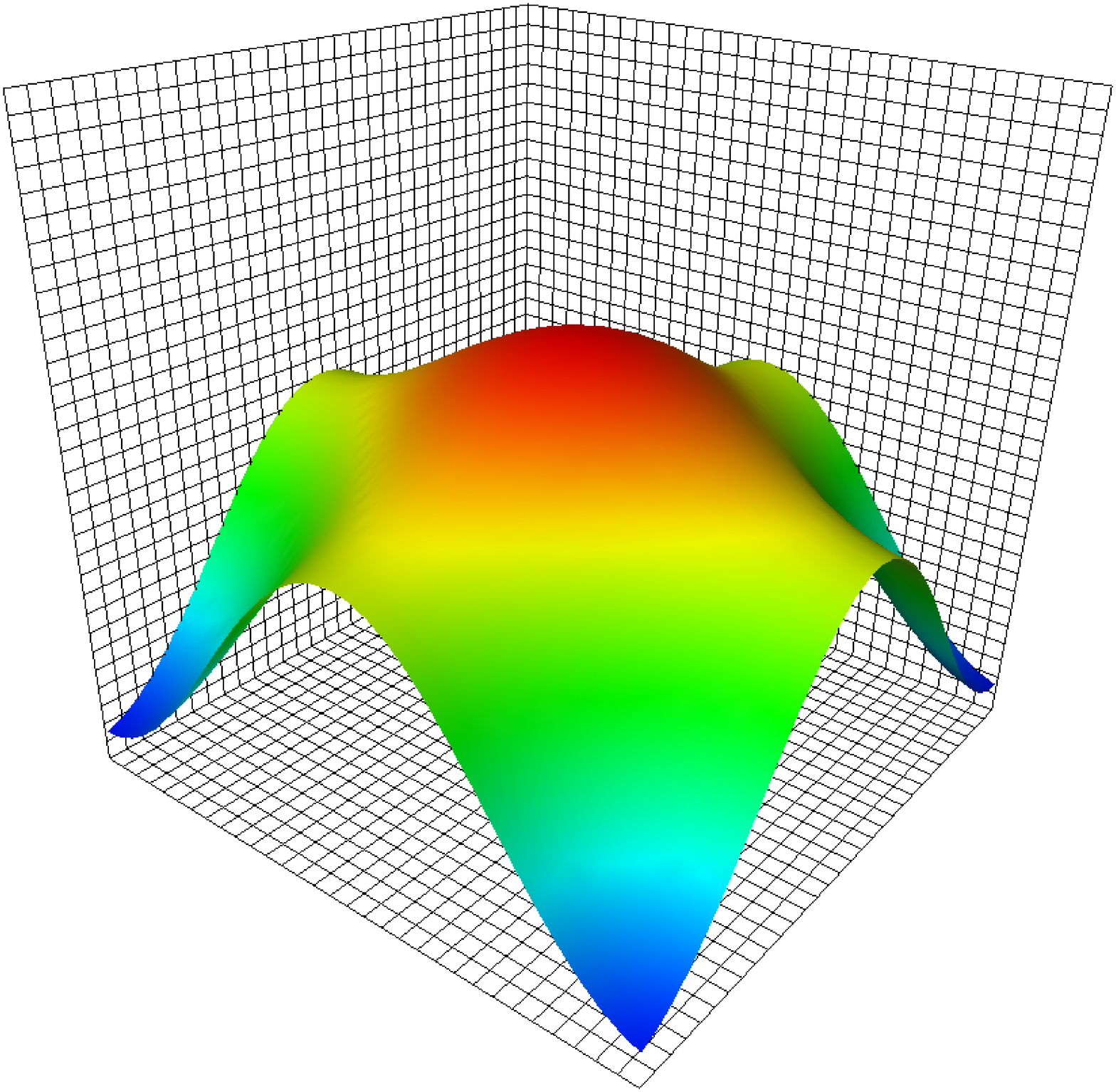}}
			{\includegraphics[width=0.24\textwidth]{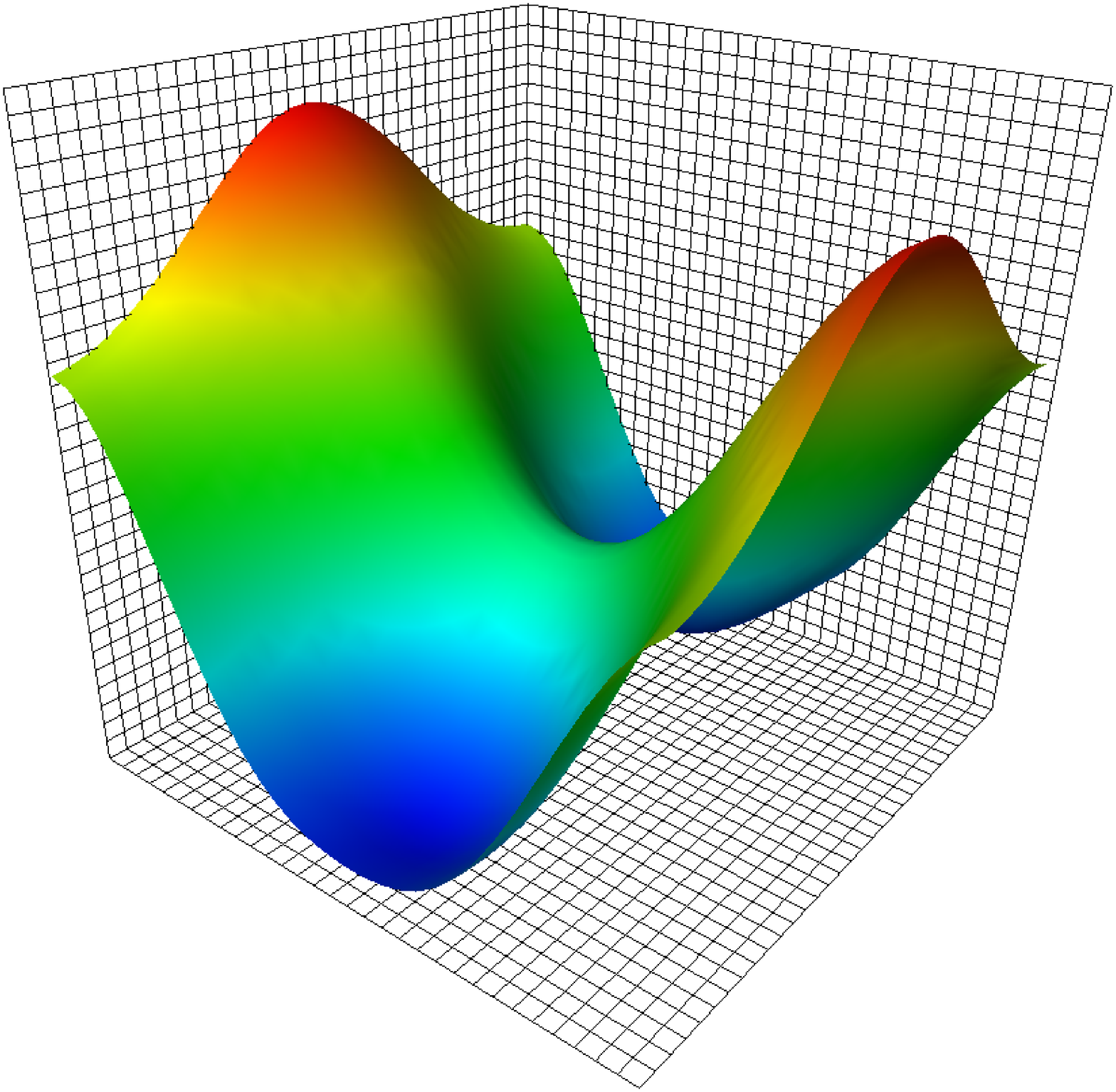}}
			{\includegraphics[width=0.24\textwidth]{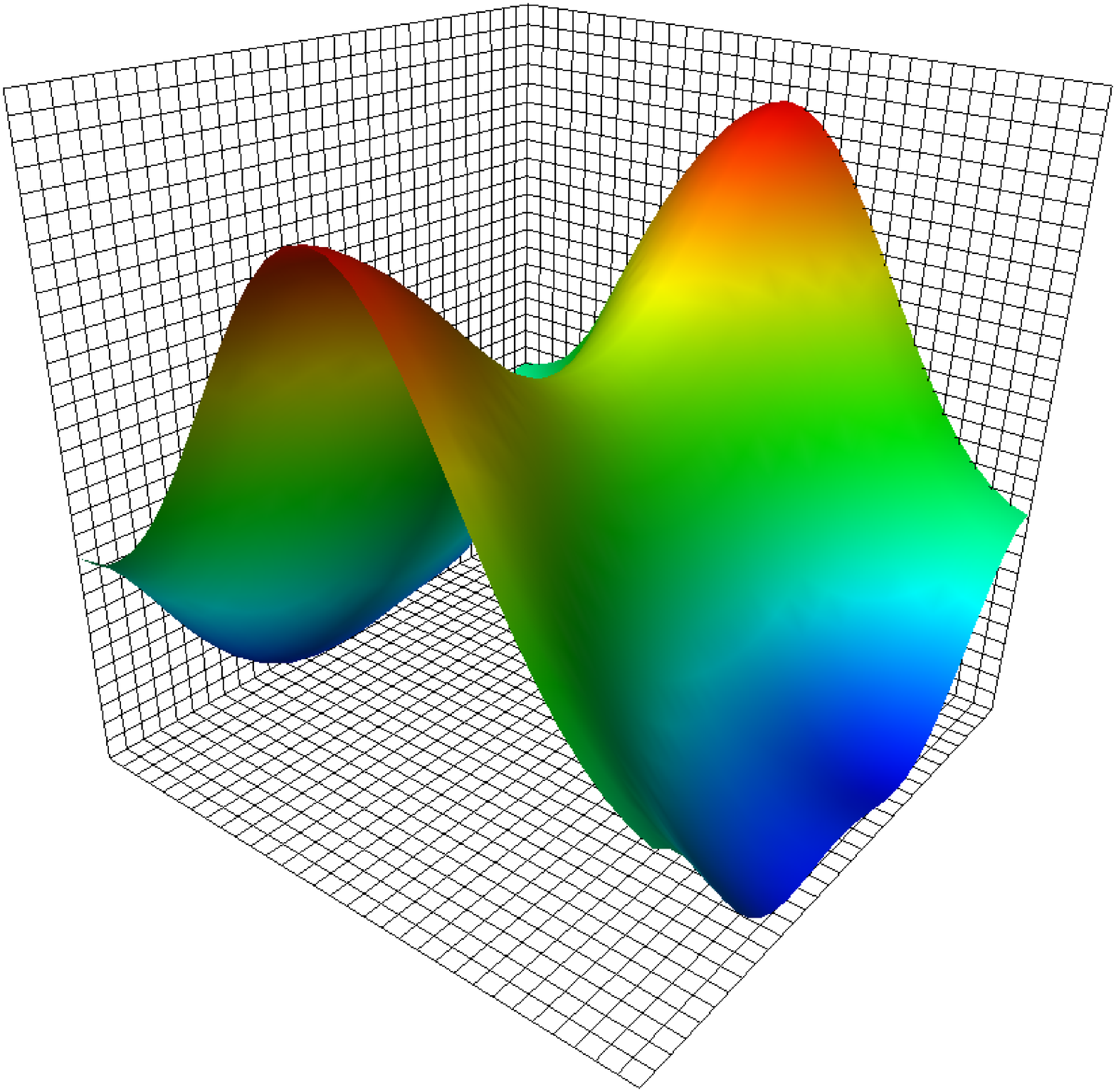}}
		\end{center}
		\caption{Lowest-lying eigenmode probability densities of the lattice
				Laplacian operator in a constant background magnetic field
				oriented in the $z$ direction are plotted as a function of the $x,y$ coordinates.
				The mode for the lowest quantised magnetic field strength relevant to
				the $d$ quark (left) is compared with the two degenerate
				eigenmodes of the second quantised magnetic field strength
				(middle, right) relevant to the $u$ quark. }
	\label{fig:emodes}
	\end{figure}

	The $qe\,B/(2M)$ term in Eq.~(\ref{dE(B)}) is a Landau level term, it corresponds to the lowest lying Landau level of the hadron.
	The Landau levels are a superposition of energy levels \cite{Bhattacharya:2007vz}
	\begin{align}
		E^2 = m^2 + {\abs{qe\,B}}\,(2\,\nu+1) - q\,\abs{e\,B}\,s + p_z^2 \nonumber, 
	\end{align}
	caused by the motion of a charged particle in an external magnetic field. Here $\nu = 0,1,2,\dots$, spin parameter $s = \pm\,1$ and $p_z$ is the component of momentum in the $\hat{z}$ direction. The charged quarks are also in an external magnetic field and in the absence of QCD would also have Landau level energies. It is possible to obtain these Landau levels using the eigenmodes of the lattice Laplacian operator for each quark in a background magnetic field. A sample of the eigenmodes for the smallest field strength are presented in Figure \ref{fig:emodes}. It is clear from Figure \ref{fig:emodes} that a particle at the centre of the Lattice will have little overlap with the Landau levels; hence knowledge of the eigenmodes may prove advantageous when constructing quark operators on the lattice.

	\par
	\subsection{Eigenmode Projections}
	Eigenmodes of the lattice Laplacian operator are calculated where no QCD effects are present; only the QED background field is present.
	\par
	Once the eigenmodes $\lambda_i$ have been obtained, the quark propagators can be projected to the eigenmodes at both the source and the sink. Projection operators $P^{n}_{QED}$ are defined
	\begin{align}
		P^{n}_{QED}(x,y) = \sum_{i=1}^{n=\abs{3\,q_f\,k_d}} \, \braket{x}{\lambda_i}\,\braket{\lambda_i}{y},
	\end{align}
	where $q_f$ is the fractional quark charge.
	The propagator is then projected at the sink using these projection operators as,
	\begin{align}
		S(x,y) = P_{QED}(x,z)\,S(z,y)
	\end{align}
	 Any combination of projection operators and smearing can be used at both the source and the sink although only a few have been investigated here.
	
	\subsection{Polarisability energy shifts}
	The energy shift due to the polarisability is smaller than that due to the magnetic moment and it also contains contributions from the Landau levels. These features make the polarisability considerably more challenging to extract than the magnetic moment. In order to extract a polarisability from the energy shifts, a relevant function must be fitted as a function of the field strength - or the field strength quanta $k_d$. However it is only sensible to do this where acceptable constant fits to the energy shift at each field strength can be adequately performed. The restrictions imposed on fitting are,
	\begin{enumerate}
		\item Constant fits to Eq.~(\ref{dE(B)}) as a function of $t$ must be acceptable;
		\item Relevant fits to Eq.~(\ref{dE(B)}) as a function of $B$ must be acceptable;
		\item Only the same fit window accross all field strengths is considered.
	\end{enumerate}
	These measures help ensure that the final fits produced are free from bias due to the selection of fit window. Best results for the neutron were found using a spatially smeared source with 100 sweeps of Gaussian smearing and a QED eigenmode projected sink.
	
	  \begin{figure}[t]
	  	\centering
	  	\begin{floatrow}
	  		\ffigbox[0.92\FBwidth]{\caption{Polarisability energy shift for all field strengths for neutron with smeared source and QED eigenmode projected sink.}\label{fig:dummy-1}}{%
	  			{\includegraphics[width=0.5\textwidth]{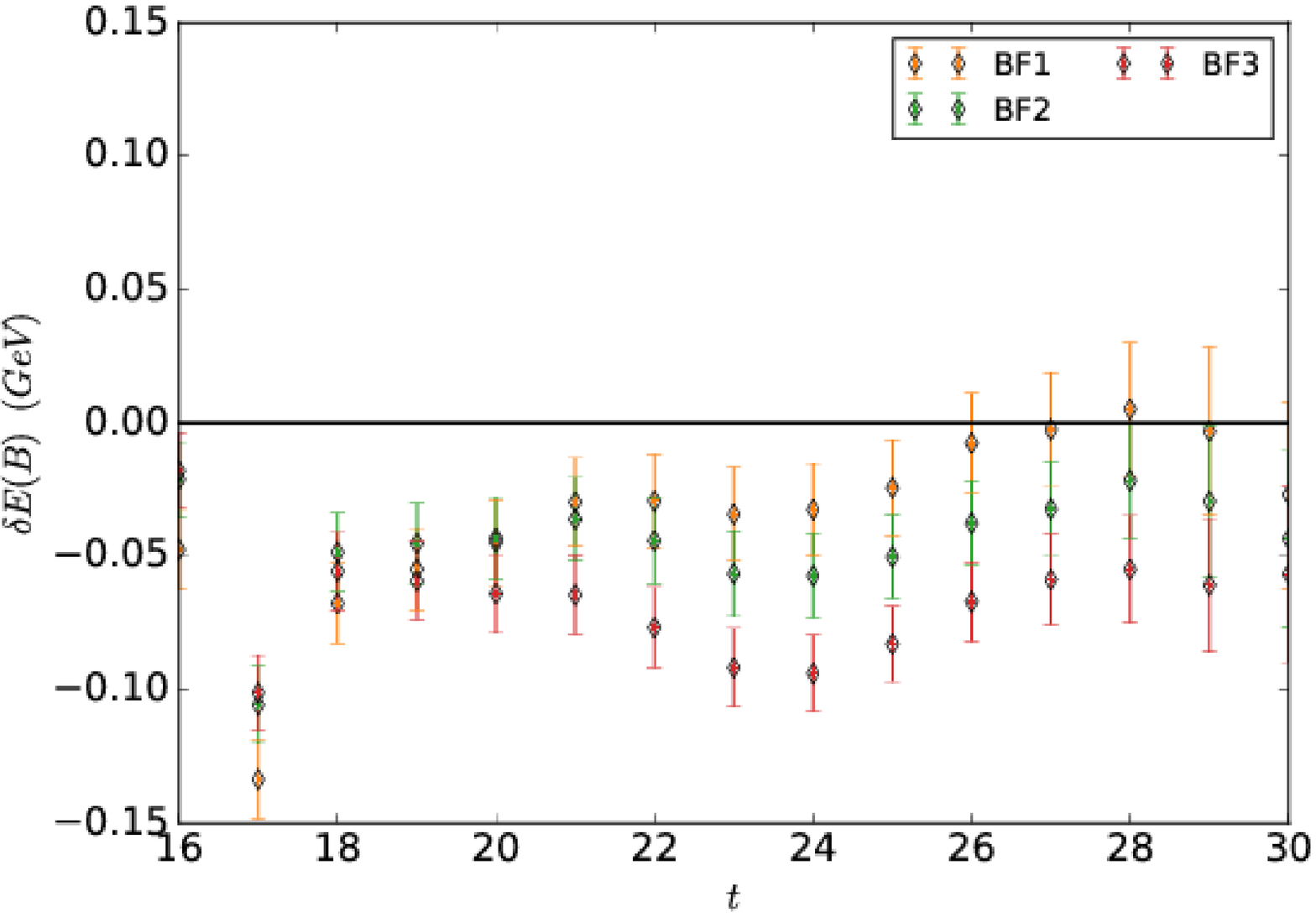}}
	  		}
	  		\ffigbox[0.92\FBwidth]{\caption{Quadratic and quadratic + linear fits of the energy shift to the field strength for the neutron with smeared source and QED eigenmode projected sink. Here $k=k_d$.}\label{fig:dummy-2}}{%
	  			{\includegraphics[width=0.5\textwidth]{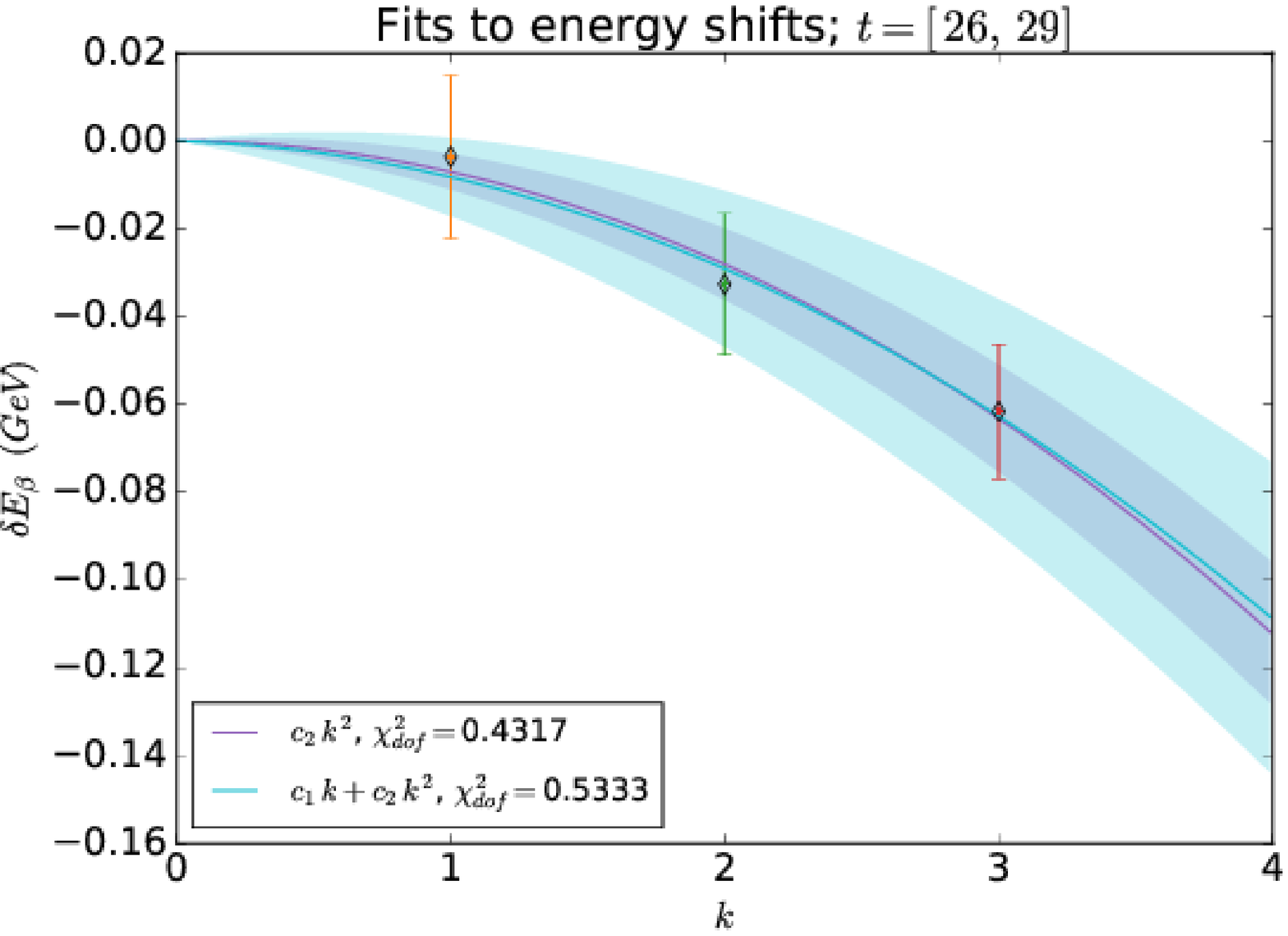}}
	  		}
	  	\end{floatrow}

	  \end{figure}

	For the neutron, it is clear from Figure \ref{fig:dummy-2} that a fit which is quadratic-only is sufficient to describe the energy shift as a function of the field strength $B$. That is, the neutron doesn't have a Landau level energy term. This is as expected as the neutron is a neutrally charged particle. 
	From the fitted curves the effective charge of the hadron and the magnetic polarisabilty can be extracted.
	The quadratic only fit results in a value of $\beta_n = 1.31(38) \times 10^{-4}$ fm$^3$ for our pion mass of $413$ MeV.
	\section{Magnetic Moment}
	The magnetic moment of a system of charged particles, $\vec{\mu}$ is related to the tendency of the system to become aligned with the external magnetic field. Returning to Eq.~(\ref{EofB}), the magnetic moment is a first order term in $B$. This results in a much larger shift in energies than the polarisability term making magnetic moments easier to extract. Of additional use in isolating the magnetic moment energy shift is the spin and field direction dependence.
	\par
	 By forming combinations of spin up and down correlation functions at both positive and negative field strengths the magnetic moment term of the energy shift can be efficiently isolated. This can be done using the ratio,
	\begin{align}
		R_m(B,t) =&  \left( \frac{G_{\downarrow}(B+,t) + 
			G_{\uparrow}(B-,t)  }{ G_{\downarrow}(0,t) + G_{\uparrow}(0,t)}
		\right) \, \left( \frac{G_{\downarrow}(0,t) + G_{\uparrow}(0,t)}{G_{\downarrow}(B-,t) + G_{\uparrow}(B+,t)} \right) 
		\nonumber\\
		&= \left( \frac{G_{\downarrow}(B+,t) + 
			G_{\uparrow}(B-,t)  }{G_{\downarrow}(B-,t) + G_{\uparrow}(B+,t) }
		\right) 
		\label{RM(B,t)}
	\end{align}
	In an analogous way to an effective mass, a magnetic moment energy shift can be found,
	\begin{align}
		\Delta E_m(B,t) &= \frac{1}{\delta t}\,\log( \frac{ R_\mu(B,t)}{R_\mu(B,t+\delta t)}) \nonumber \\
		&= -\mu\,B
		\label{dEmB}
	\end{align}
	This formulation of the energy shift has the advantage that it removes many of the correlated errors between spin orientations.
	\par
	In the same manner as the magnetic polarisability, magnetic moment energy shifts have constant plateaus fitted to them. This time linear or linear $+$ cubic terms are considered. This cubic term is appropriate as it corresponds to the next lowest order term in Eq.~(\ref{dEmB}).
	
	  \begin{figure}[t]
	  	\centering
	  	\begin{floatrow}
	  		\ffigbox[0.92\FBwidth]{\caption{Magnetic moment energy shift for the three field strengths considered for the neutron with a smeared source and a QED-eigenmode projected sink.}\label{fig:neutron-1-m}}{%
	  			{\includegraphics[width=0.5\textwidth]{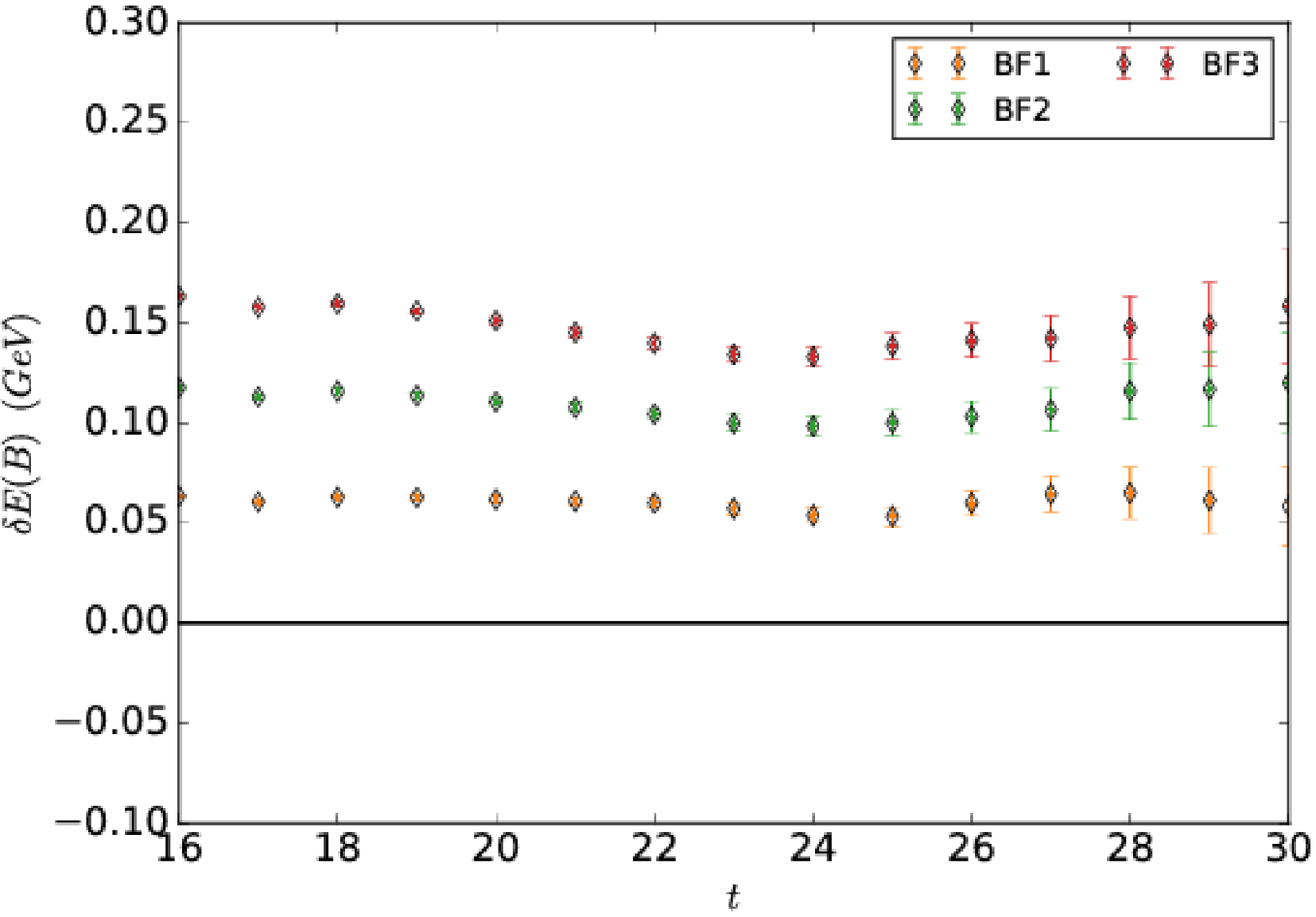}}
	  		}
	  		\ffigbox[0.92\FBwidth]{\caption{Linear and linear $+$ cubic fits of the energy shift to the field strength for the neutron with a smeared source and a QED-eigenmode projected sink.}\label{fig:neutron-2-m}}{%
				{\includegraphics[width=0.5\textwidth]{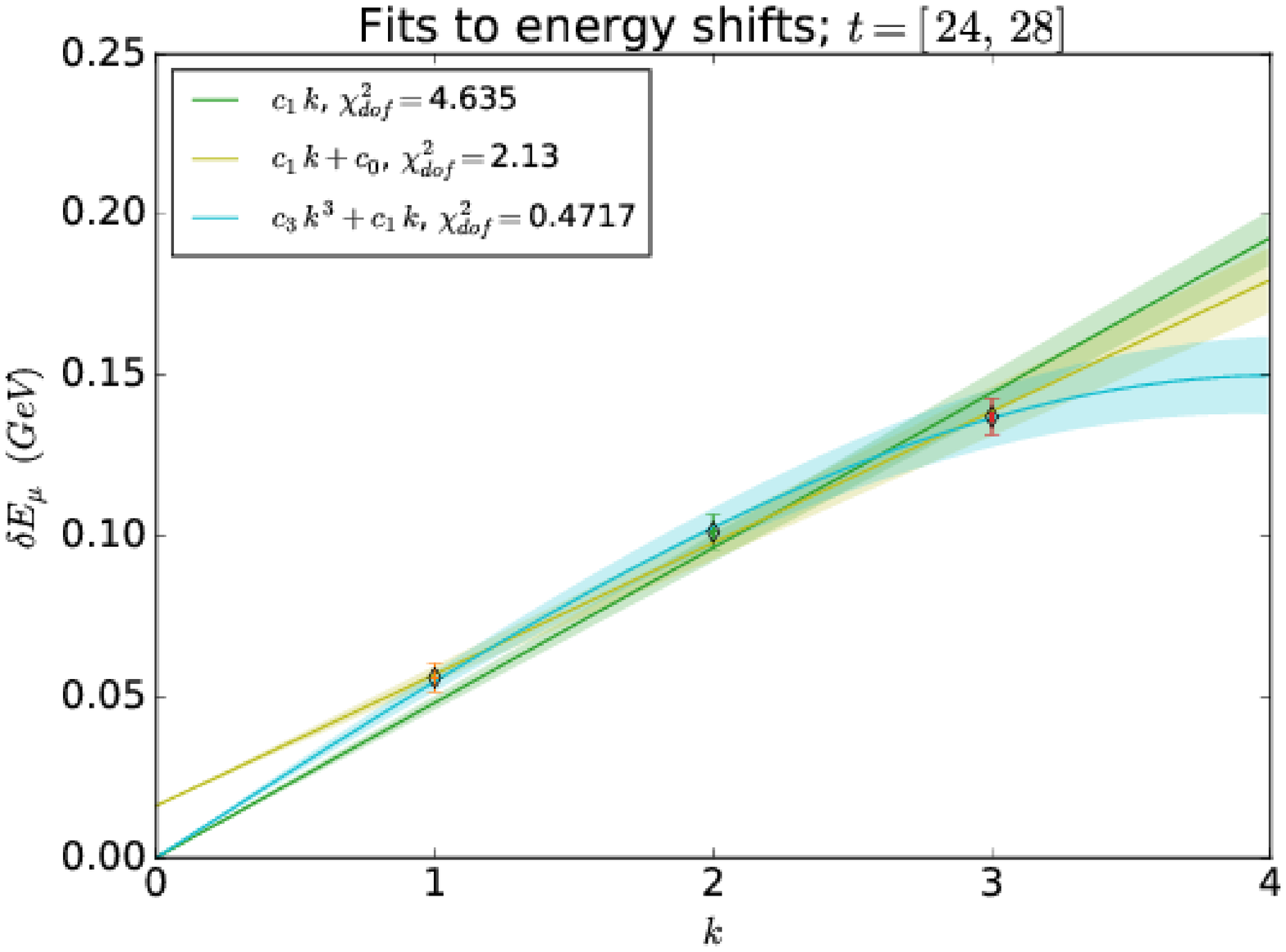}}
	  		}
	  	\end{floatrow}	
	
	\end{figure}
	
	  \begin{figure}[t]
	  	\centering
	  	\begin{floatrow}
	  		\ffigbox[0.92\FBwidth]{\caption{Magnetic moment energy shift for the three field strengths considered for proton with smeared source and QED eigenmode projected sink.}\label{fig:proton-1-m}}{%
	  			{\includegraphics[width=0.5\textwidth]{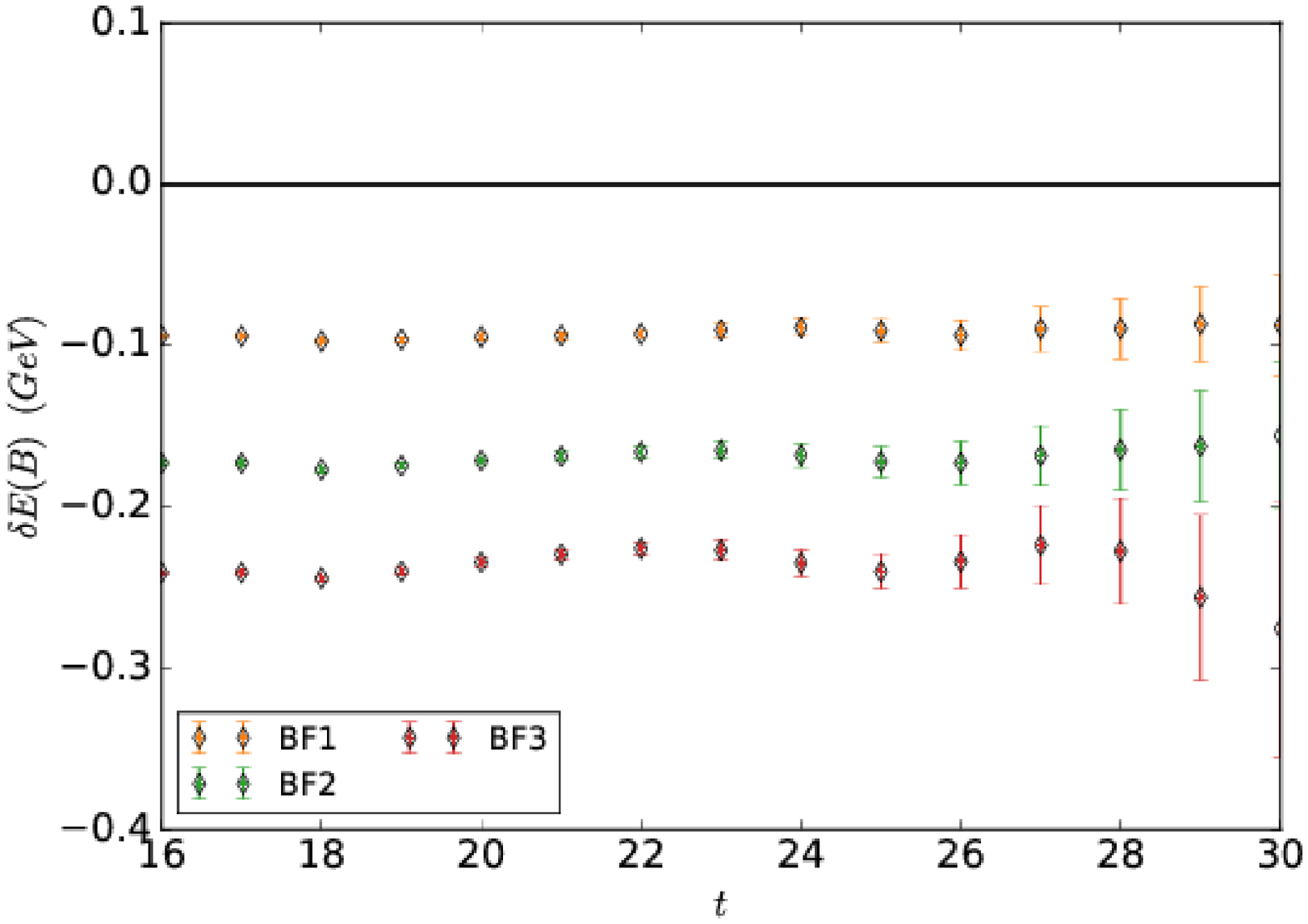}}
	  		}
	  		\ffigbox[0.92\FBwidth]{\caption{Linear and linear $+$ cubic fits of the energy shift to the field strength for the proton with smeared source and QED eigenmode projected sink.}\label{fig:proton-2-m}}{%
	  			{\includegraphics[width=0.5\textwidth]{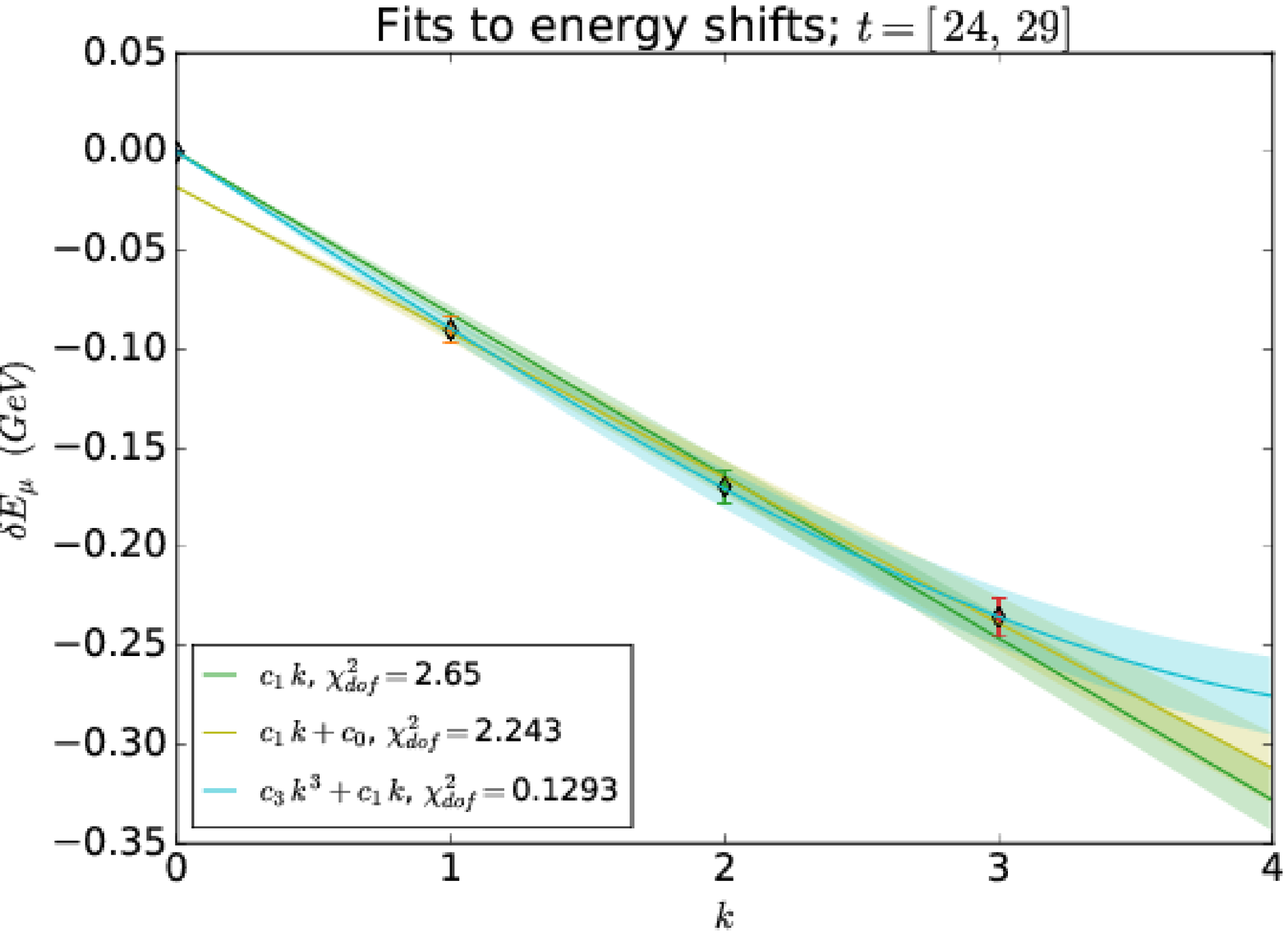}}
	  		}
	  	\end{floatrow}
	  	
	  \end{figure}	
	
	From Figures \ref{fig:neutron-2-m} and \ref{fig:proton-2-m} it is clear that the cubic term is necessary in order to adequately fit the energy shifts. This suggests that the third field strength in particular is becoming too large for the energy relation in Eq.~(\ref{EofB}) to fully describe the system. This could be remedied by using a larger lattice volume and corresponding smaller field strengths.
	\par
	The magnetic moments for the proton and neutron linear $+$ cubic fits are shown in Table \ref{mu}. Good agreement is seen with results from the alternative three-point function method on the same lattices and at the same pion mass\cite{BenOwen}.
	\begin{table}[!ht]
		\centering
		\caption{Proton and neutron magnetic moments at a pion mass of $413$ MeV}	
		\begin{tabular}{lll}
			\hline \hline
			& \phantom{0-}This Work  & \phantom{0-}3PT Method \\ \hline 
			\multicolumn{1}{l}{proton}  &\phantom{-0}$2.20(16)\,\mu_N$ & \phantom{0-}$2.184(22)\,\mu_N$\\
			\multicolumn{1}{l}{neutron} &$-1.36(10)\,\mu_N$ & $-1.371(14)\,\mu_N$ \\
			\hline \hline
		\end{tabular}%
		\label{mu}	
	\end{table}		
	\newpage
	\section{Conclusion}
	Through the use of Landau eigenmode projectors in the sinks of the quark propagators, we have been able to observe plateaus in the correlation functions describing the magnetic polarisability for the first time. We have also examined the utility of a Landau level projector for the proton in its final state. The results are encouraging and the refinement of creation and annihilation operators is in progress. Efforts to expand this method to excited and negative parity states are desirable as well as chiral extrapolations to enable confrontation of experiment. The background field method has been shown again to be a useful tool to access magnetic properties on the lattice.
	
	\acknowledgments
	This research is supported by an Australian Government Research Training Program Scholarship. This work was supported with supercomputing resources provided by the Phoenix HPC service at the University of Adelaide. This research was undertaken with the assistance of resources from the National Computational Infrastructure (NCI), which is supported by the Australian Government.

	\bibliographystyle{JHEP_arXiv}
	\bibliography{ms}

\end{document}